\documentclass{PoS}

\usepackage[utf8x]{inputenc}
\usepackage{mciteplus}
\usepackage{bm}

\newcommand{\rt}{\mathbf{r}}

\newcommand{\Deltat}{\boldsymbol{\Delta}}

\newcommand{\xt}{\mathbf{x}}
\newcommand{\yt}{\mathbf{y}}
\newcommand{\bt}{\mathbf{b}}
\newcommand{\pt}{\mathbf{p}}
\newcommand{\Pt}{\mathbf{P}}

\newcommand{\der}{\mathrm{d}}
\newcommand{\gev}{\mathrm{GeV}}

\author{\speaker{Heikki M\"antysaari} \\
Department of Physics, University of Jyväskylä, P.O. Box 35, 40014 University of Jyväskylä, Finland, and \\
Helsinki Institute of Physics, P.O. Box 64, 00014 University of Helsinki, Finland\\
E-mail: \email{heikki.mantysaari@jyu.fi} 
}

\author{Niklas Mueller \\
Physics Department, Brookhaven National Laboratory, Bldg. 510A, Upton, NY 11973, USA \\
E-mail: \email{nmueller@bnl.gov}
}
\author{Bj\"orn Schenke \\
Physics Department, Brookhaven National Laboratory, Bldg. 510A, Upton, NY 11973, USA
E-mail: \email{bschenke@bnl.gov}
}

\title{Diffractive dijet production from the Color Glass Condensate and the small-$x$ gluon distributions}

\ShortTitle{Diffractive dijet from CGC}

\abstract{We study exclusive dijet production in electron-proton deep inelastic scattering at a future Electron Ion Collider. We predict the elliptic modulation of the cross section as a function of the angle between the dijet transverse momentum and the recoil momentum, and show that this modulation  is due to non-trivial angular correlations  between the transverse coordinate and transverse momentum in the Wigner (or Husimi) distribution. The small-$x$ evolution is shown to decrease the elliptic modulation in the EIC kinematics,  because of the growth of the proton with decreasing $x$.}

\FullConference{XXVII International Workshop on Deep-Inelastic Scattering and Related Subjects - DIS2019\\
		8-12 April, 2019\\
		Torino, Italy}

\begin{document}

\section{Introduction}

The structure of the proton is a result of complicated non-perturbative many-body interactions between its fundamental building blocks, the quarks and gluons. This structure is encoded in various distribution functions, the simplest one being the collinear parton distribution functions that describe the parton density as a function of longitudinal momentum fraction $x$ carried by the parton (measured at a given scale $Q^2$). These distributions have been measured with great precision by experiments at HERA~\cite{Aaron:2009aa} by studying total electron-proton cross sections.

More detailed information can be extracted from more differential observables that provide access to more differential distribution functions. For example, in exclusive photon or vector meson production the total momentum transfer is measurable, and via Fourier transform provides access to the spatial distribution of partons known as the generalized parton distribution function GPDF~\cite{Diehl:2003ny,Belitsky:2005qn}. It is also possible to study the distribution of partons in the proton in transverse momentum space, described by transverse momentum dependent parton distribution functions (TMDs)~\cite{Collins:1981uw}.

The most complete information of the proton structure is encoded in the Wigner distribution~\cite{Ji:2003ak,Belitsky:2003nz,Lorce:2011kd}, which depends on both transverse coordinate and transverse momentum, in addition to the longitudinal momentum fraction $x$. This quantum distribution is not positive definite and has a probabilistic interpretation only in certain semi-classical limits \cite{Moyal:1949sk,Hillery:1983ms,Polkovnikov:2009ys}.
To access the Wigner distribution, more differential observables than single particle production or total cross sections are needed. In Ref.~\cite{Hatta:2016dxp}, it was shown that diffractive dijet production, where two jets are produced in a process where no net color charge is exchanged with the target, is sensitive to the gluon Wigner distribution at small $x$. A future Electron Ion Collider in the US~\cite{Accardi:2012qut,Aschenauer:2017jsk} or LHeC~\cite{AbelleiraFernandez:2012cc} at CERN would be able to measure this process over a wide kinematical region at high center-of-mass energies.

At high energies or small $x$ the convenient effective theory to describe high energy scattering processes is provided by the Color Glass Condensate (CGC), which describes Quantum Chromodynamics (QCD) in the high energy limit. In Ref.~\cite{Mantysaari:2019csc}, summarized here, we calculate both the diffractive dijet production cross section and the Wigner distribution in the CGC framework.

%  \cite{JalilianMarian:1996xn,JalilianMarian:1997jx,JalilianMarian:1997gr,Iancu:2001md,Ferreiro:2001qy,Iancu:2001ad,Iancu:2000hn}

\section{Dipole-proton interaction in the CGC}
\label{sec:dipole}
At high energies the convenient degrees of freedom are Wilson lines $U(\xt)$ that describe the color rotation that the parton encounters when propagating eikonally through the target. For a given target configuration, the Wilson lines are obtained by solving the Yang-Mills equations
\begin{equation}
U(\xt) = P \exp \left( -ig \int \der x^- \frac{\rho (x^-,\xt)}{\nabla^2 + \tilde m^2 }\right).
\end{equation}
The color charge density $\rho$ is assumed to be a local Gaussian variable with expectation value being related to the local density of the proton, which we assume to be Gaussian in this work, and $\tilde m^2$ is an infrared regulator. After the Wilson lines are determined at the initial Bjorken-$x$, their evolution to smaller $x$ is obtained by solving the perturbative JIMWLK evolution equations (see e.g.~\cite{JalilianMarian:1996xn}).
%JalilianMarian:1997jx,JalilianMarian:1997gr,Iancu:2001md,Ferreiro:2001qy,Iancu:2001ad,Iancu:2000hn}.
  All parameters that control e.g. the density at the initial $x_0=0.01$, the size of the proton and the values of the strong coupling and infrared regulators are constrained by the HERA structure function and diffractive vector meson production measurements~\cite{Mantysaari:2018zdd}. 
%For example, the JIMWLK kernel $K^i(\xt)$ describing the gluon emission in coordinate space, is regulated by an effective mass $m$ to exponentially suppress contributions at non-perturbatively large distance scales by replacing 
%\begin{equation}
%	K^i(\xt) = \frac{x^i}{\xt^2} \rightarrow m |\xt| K_1(m|\xt|)  \frac{x^i}{\xt^2}.
%\end{equation}
  For a more detailed description of the setup, the reader is referred to~\cite{Mantysaari:2019csc}.

When Wilson lines are sampled on the lattice and evolved to smaller $x$ with the JIMWLK equation, it becomes possible to construct the dipole-target scattering amplitude at any $x$
\begin{equation}
N\left( \rt = \xt - \yt, \bt = \frac{\xt + \yt}{2} \right) = 1 - \frac{1}{N_c} \langle \mathrm{Tr} U(\xt) U^\dagger(\yt) \rangle,
\end{equation}
where the average is taken over different possible target configurations.

\begin{figure}[tb]
    \centering
    \begin{minipage}{.48\textwidth}
        \centering
        \includegraphics[width=\textwidth]{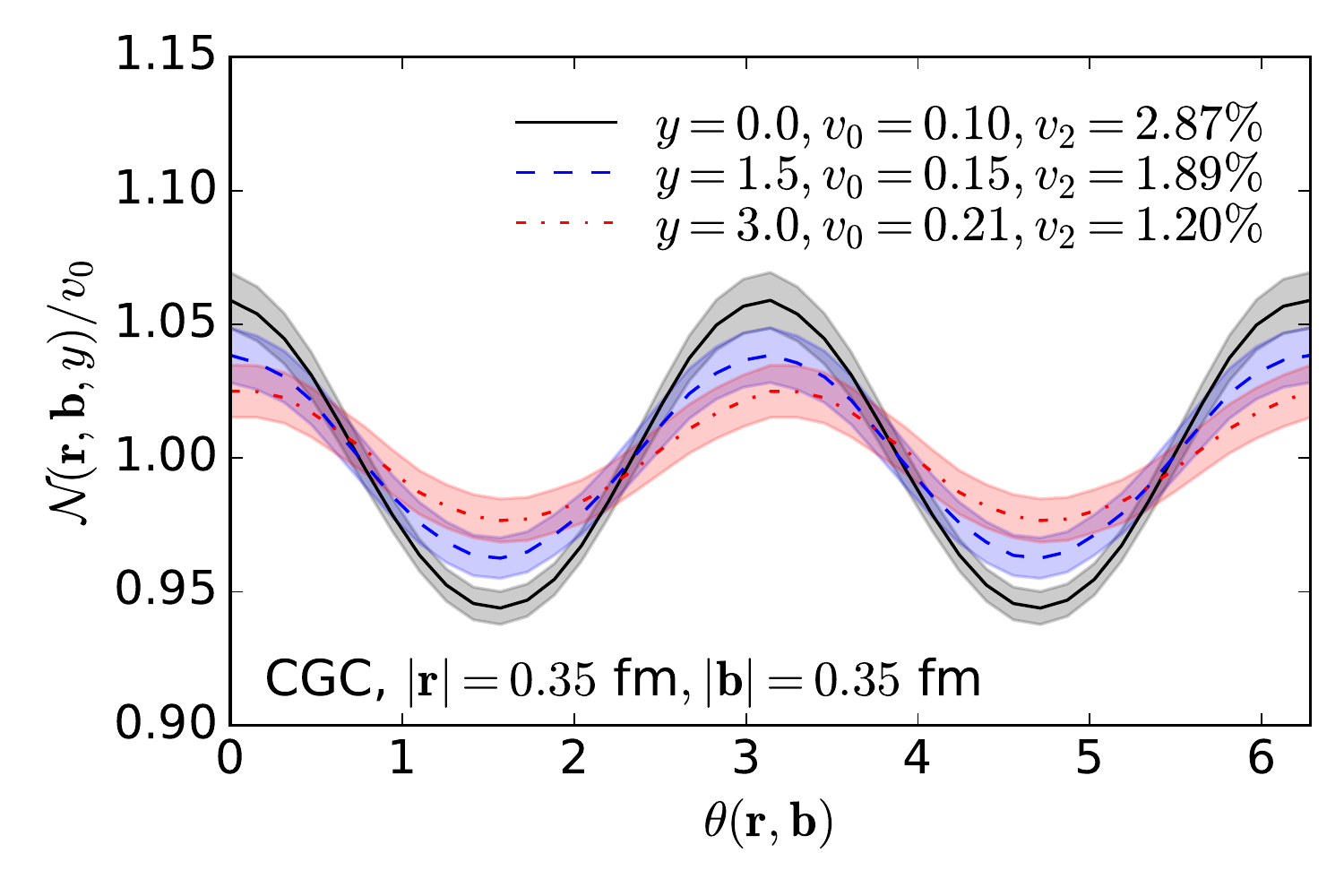} 
				\caption{Dipole amplitude as a function of angle between dipole size and impact parameter at different rapidities. Figure from Ref.~\cite{Mantysaari:2019csc}.}
				%as a function of the number of charged hadrons per pseudo-rapidity interval around mid-rapidity compared }
		\label{fig:dipole}
    \end{minipage} \quad
    \begin{minipage}{0.48\textwidth}
        \centering
      \includegraphics[width=\textwidth]{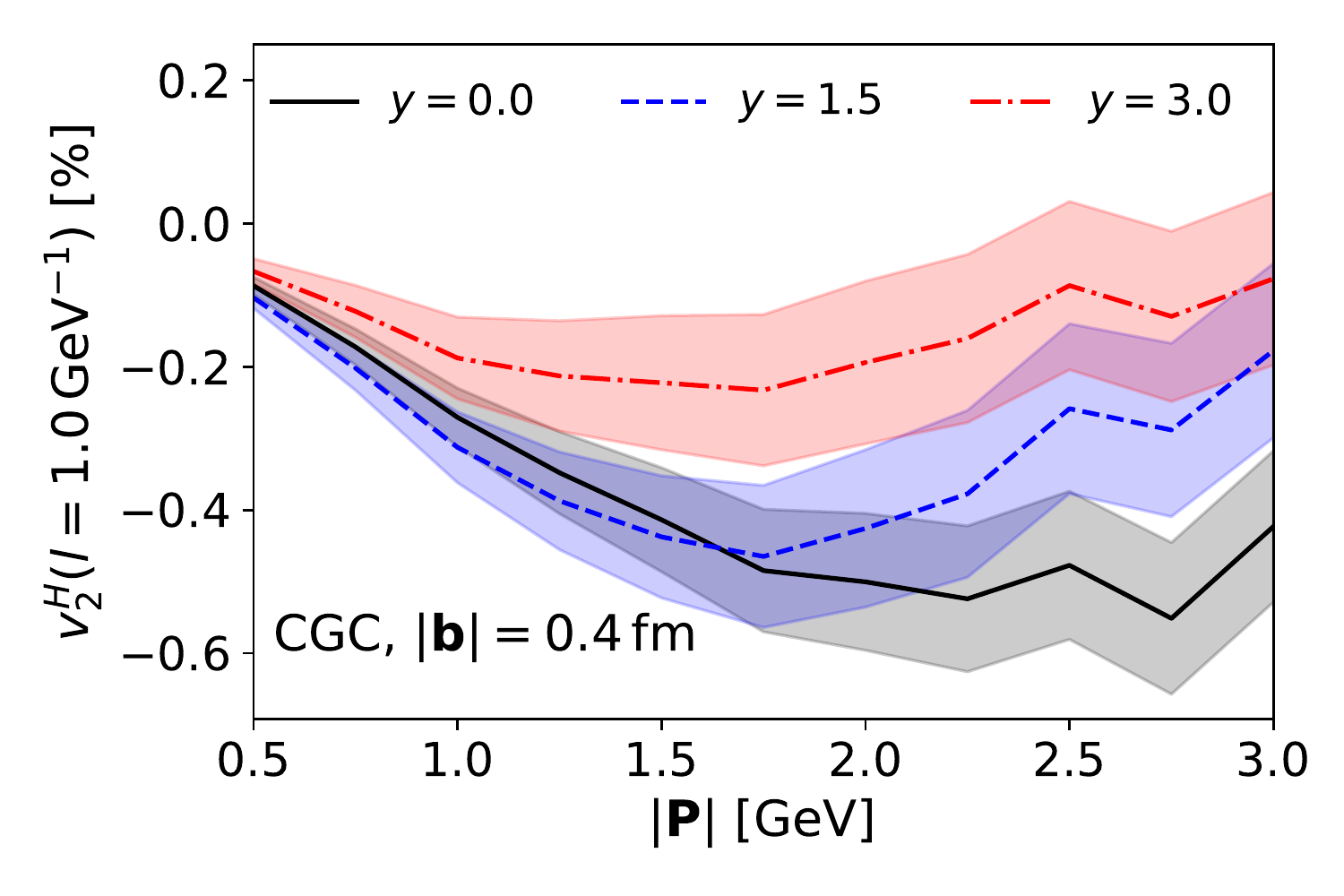} 
				\caption{Rapidity evolution of the elliptic component of the Husimi distribution from Ref.~\cite{Mantysaari:2019csc}.}
		\label{fig:husimi}
    \end{minipage}
\end{figure}

When we consider exclusive dijet production, the Fourier conjugates to the dijet momentum and to the recoil momentum are the dipole size $\rt$ and  impact parameter $\bt$. Having this in mind, we study the angular modulation of the dipole-proton scattering amplitude $N(\rt,\bt)$ calculated from the CGC framework. The dipole amplitude $N$ as a function of the angle between $\rt$ and $\bt$ is shown in Fig.~\ref{fig:dipole}. Note that in widely used dipole amplitude parametrizations such as IPsat~\cite{Kowalski:2003hm} there would be no angular dependence. 

To quantify the evolution of the elliptic modulation of the dipole amplitude, we extract the Fourier harmonics $v_n$ writing the dipole amplitude as
%\begin{equation}
$	N(\rt,\bt) = v_0 [1 + 2 v_2 \cos 2\theta(\rt,\bt) ]$.
%\end{equation}
The extracted $v_0$ and $v_2$ coefficients at different rapidities are shown in Fig.~\ref{fig:dipole}. We find that the evolution suppresses the elliptic modulation (note that Bjorken-$x$ is related to the evolution rapidity as $x = x_0 e^{-y}$ with $x_0=0.01$). This is mainly due to the rapid growth of the proton density in the dilute region, resulting in a smoother and larger proton with smaller density gradients

\section{Wigner and Husimi distributions}
As discussed in the Introduction, the gluon Wigner distribution $xW(\Pt, \bt, x)$ contains the most  complete information of the small-$x$ gluonic structure of the proton. In particular, it describes the gluon distribution as a function of both transverse coordinate $\bt$ and transverse momentum $\Pt$. The disadvantage is that due to the uncertainty principle it can not have a probabilistic interpretation, and we indeed show in Ref.~\cite{Mantysaari:2019csc} that when calculated from the CGC framework, the Wigner distribution becomes negative at small transverse momenta.

If the Wigner distribution is smeared over both transverse coordinate and transverse momentum with the smearing parameters being inverse to each other, one obtains the so called Husimi distribution
\begin{equation}
xH(\Pt,\bt,x)) = \frac{1}{\pi^2} \int \der^2 \bt' \der^2 \Pt' e^{-\frac{1}{l^2}(\bt-\bt')^2 - l^2(\Pt-\Pt')^2} xW(\Pt',\bt',x). 
\end{equation}
In this work we choose $l=1\,\mathrm{GeV}^{-1}$, as it corresponds to a distance scale much smaller than the proton size, but does not result in too large smearing in momentum space that would wash out most of the transverse momentum dependence. In Ref.~\cite{Mantysaari:2019csc} it is shown that the Husimi and Wigner distributions agree at large $|\Pt| \gtrsim 1/l$, and that the Husimi distribution calculated from the CGC framework following~\cite{Hagiwara:2016kam} is positive definite. 

To study the elliptic modulation (dependence on the angle between $\Pt$ and $\bt$), we write the Husimi distribution as
\begin{equation}
xH(\Pt,\bt,x) = v_0^H [1 + 2 v_2^H \cos 2\theta(\Pt,\bt)].
\end{equation}
The rapidity evolution of the elliptic coefficient $v_2^H$  is shown in Fig.~\ref{fig:husimi}. Except at the smallest momenta, the evolution suppresses the elliptic modulation as expected based on the analysis of the dipole amplitude in Sec.~\ref{sec:dipole}. At the smallest momentum values the elliptic component first grows in the evolution, which can be understood as when the proton grows, large dipoles $|\rt| \sim |\Pt|^{-1}$ with large elliptic modulation start to contribute, before the proton grows enough that we start to see the decreasing density gradients also at small $|\Pt|$.

\section{Diffractive dijet production}
As discussed in Ref.~\cite{Hatta:2016dxp}, diffractive dijet production is sensitive to the gluon Wigner distribution. In particular, it is interesting to study diffractive dijet production as a function of the two momentum vectors, the average momentum $\Pt = \frac{1}{2}(\pt_1 - \pt_2)$ and the recoil momentum $\Deltat = \pt_1 + \pt_2$, where $\pt_1$ and $\pt_2$ are the momenta of the individual jets. The connection to the Wigner distribution is shown in the correlation limit $|\Pt| \gg |\Deltat|$, thus we use $|\Deltat|=0.1$ GeV in this work. 

\begin{figure}[tb]
    \centering
    \begin{minipage}{.48\textwidth}
        \centering
        \includegraphics[width=\textwidth]{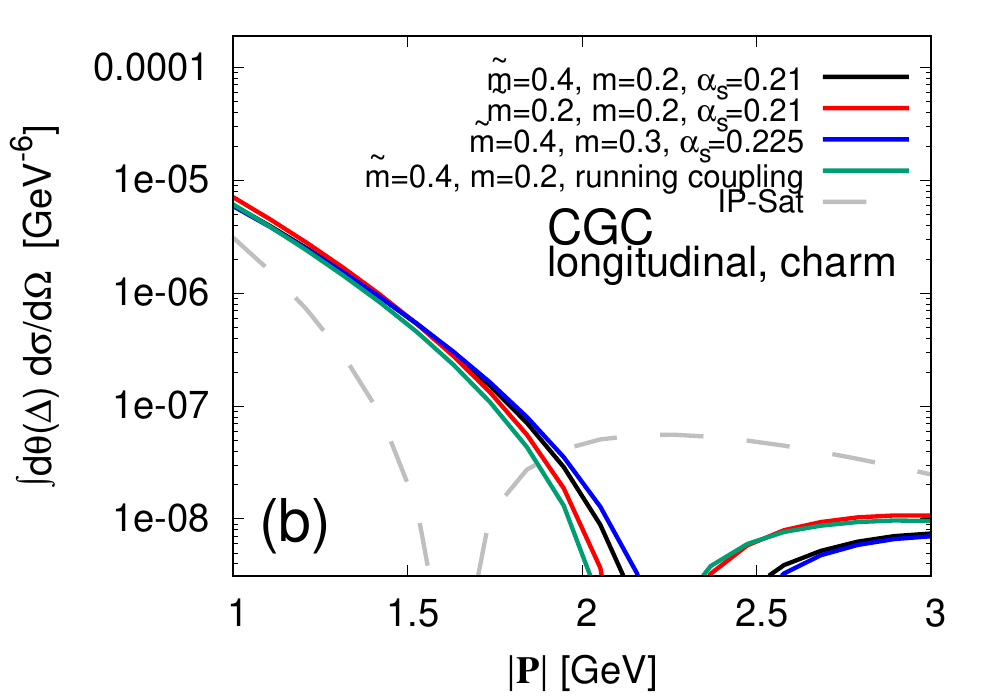} 
				\caption{Charmed dijet photoproduction spectrum as a function of dijet momentum. Figure from Ref.~\cite{Mantysaari:2019csc}.}
		\label{fig:dijet_spectrum}
    \end{minipage} \quad
    \begin{minipage}{0.48\textwidth}
        \centering
      \includegraphics[width=\textwidth]{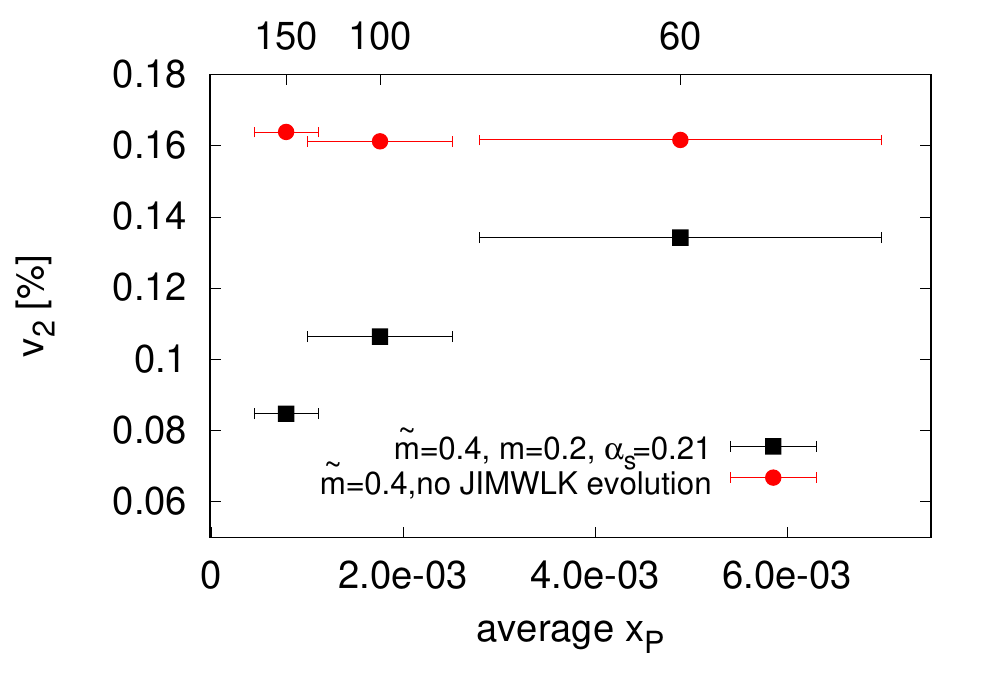} 
				\caption{Elliptic ($v_2$) modulation of dijet photoproduction as a function of momentum fraction $x_p$ of the target. The values on top refer to the  center of mass energies $W$. Figure from Ref.~\cite{Mantysaari:2019csc}.}
		\label{fig:dijet_v2}
    \end{minipage}
\end{figure}

The diffractive dijet production cross section in the CGC framework is derived in Ref.~\cite{Altinoluk:2015dpi}. The cross section as a function of jet momentum $\Pt$ is shown in Fig.~\ref{fig:dijet_spectrum}, where results obtained with different infrared regulators (with value of the coupling constant adjusted to describe the HERA structure function data) and using both fixed and running coupling evolution are shown, and our results are found to be insensitive to the infrared regularization. Here it is worth noticing that the Fourier conjugate to $\Pt$ is the dipole size, and thus this can be seen as a diffraction off the $q\bar q$ Fock state of the probing photon. Here we only consider charmed dijets, so the diffractive dip location can be estimated to be $|\rt_\gamma|^{-1} \sim \sqrt{m_c^2 + Q^2} \approx 1.4\,\gev$.

To study the elliptic modulation we extract the Fourier harmonics of the dijet production cross section:
\begin{equation}
\der\sigma = v_0[1 + 2 v_2 \cos 2\theta(\Pt,\Deltat)].
\end{equation}
The elliptic coefficient $v_2$ as a function of Bjorken-$x$ (denoted as $x_p$)  is shown in Fig.~\ref{fig:dijet_v2}, where we find that the energy evolution reduces $v_2$ by almost a factor of $2$ in the EIC energy range. This is mostly due to the increasing proton size suppressing density gradients. The modulation is relatively small and likely difficult to measure. However, at larger $|\Deltat|$ we expect a much larger signal~\cite{Salazar:2019ncp}. For comparison, we also show the result obtained in the case where we do not perform the JIMWLK evolution towards small-$x$, but just scale the overall proton density, in which case the $v_2$ is independent of energy. In the models where there is no dependence on the angle between $\rt$ and $\bt$ in the dipole amplitude, one gets exactly $v_2=0$~\cite{Altinoluk:2015dpi}.

\section{Conclusions}

We have calculated the Wigner and Husimi distributions from the CGC framework and the diffractive dijet production cross section, which in principle is sensitive to the gluon Wigner distribution at small $x$. 
By solving the perturbative JIMWLK evolution equations we find that in the gluon distributions the correlation between the momentum and coordinate space angles decreases when evolving towards small-$x$, and predict that the elliptic modulation in the dijet production cross section decreases almost by a factor of $2$ from the lowest to highest center of mass  energies at the EIC.

%\bps{rewrite last sentence. not sure what you want to say} We predict small elliptic modulations in the diffractive charmed dijet production at the EIC in the kinematics where the connection to the Wigner distribution is most rigorous.

\subsection*{Acknowledgements}
HM is supported by the Academy of Finland, project 314764, and by the European Research Council, Grant ERC-2015-CoG-681707. NM and BS are supported by the U.S. Department of Energy, Office of Science, Office of Nuclear Physics, under contract No.DE- SC0012704. NM is funded by the Deutsche Forschungsgemeinschaft (DFG, German Research Foundation) - Project 404640738.
	
\bibliographystyle{JHEP} 
\bibliography{refs}
\end{document}